# Conveying Situational Information to People with Visual Impairments

**Tousif Ahmed**
Indiana University Bloomington
Bloomington, USA
touahmed@indiana.edu

**Rakibul Hasan**
Indiana University Bloomington
Bloomington, USA
rakhasan@indiana.edu

**Kay Connelly**
Indiana University Bloomington
Bloomington, USA
connelly@indiana.edu

**David Crandall**
Indiana University Bloomington
Bloomington, USA
djcran@indiana.edu

**Apu Kapadia**
Indiana University Bloomington
Bloomington, USA
kapadia@indiana.edu

## ABSTRACT

Knowing who is in one's vicinity is key to managing privacy in everyday environments, but is challenging for people with visual impairments. Wearable cameras and other sensors may be able to *detect* such information, but how should this complex visually-derived information be *conveyed* in a way that is discreet, intuitive, and unobtrusive? Motivated by previous studies on the specific information that visually impaired people would like to have about their surroundings, we created three medium-fidelity prototypes: 1) a 3D printed model of a watch to convey tactile information; 2) a smartwatch app for haptic feedback; and 3) a smartphone app for audio feedback. A usability study with 14 participants with visual impairments identified a range of practical issues (e.g., speed of conveying information) and design considerations (e.g., configurable privacy bubble) for conveying privacy feedback in real-world contexts.

**INTRODUCTION**

People with visual impairments may find it challenging to maintain situational awareness about the social environment around them. Knowing if people are nearby is very important in everyday environments, and not knowing may create privacy [2], security [4, 15], and safety [3, 5] risks. To avoid these risks, people with visual impairments may avoid using mobile and computing devices in public due to fear of eavesdropping [1, 2], which introduces situationally-induced impairments and disabilities (SIIDs). Now that modern mobile and wearable cameras can be combined with powerful cloud-based computer vision services (e.g., IBM visual recognition[1], Google Cloud Vision. [2]), it is becoming feasible to automatically sense properties of the social environment. Microsoft's Seeing AI project[3], for example, recently implemented an iPhone application that can describe the people nearby and estimate their distance from the camera. However, while the computer vision challenges are being addressed, it is unclear how to actually communicate complex information sensed about the environment to people in an efficient and unobtrusive but non-visual way.

Relaying the number of people and their proximity is complicated for multiple reasons. First, our social surroundings are extremely dynamic, changing moment to moment as people move around us and as we move through groups of people. Practical feedback mechanisms need to provide this in a way that does not overwhelm the user. Second, feedback must not interfere with the user's other senses, since many people with visual impairments use their hearing as well as other devices to monitor the environment. Third, information needs to be conveyed quickly and discreetly, delivering feedback that is timely while not attracting undue attention that might exacerbate privacy and safety risks. For these reasons, obvious solutions [9, 10] like verbally describing the surroundings through headphones may not work well in practice. Some previous work [6, 7] has explored this problem and designed a haptic belt [7] to convey the information in an unobtrusive manner. However, this solution requires custom hardware, which may limit its adoption and social acceptability [12]. We focus on wrist-worn devices like Smartwatches and fitness trackers, which are becoming ubiquitous [4].

To study efficient and effective ways to convey privacy-related information about the environment through a wearable device, we implemented three different modes of feedback prototypes on wrist-worn devices. This extended abstract presents the prototypes and the findings of our exploratory study with 14 participants with visual impairments. Our discussions with the participants provide useful design considerations for providing privacy feedback.

---

[1]IBM Visual Recognition. https://www.ibm.com/watson/services/visual-recognition/

[2]Google Cloud Vision. https://cloud.google.com/vision/

[3]Seeing AI. https://www.microsoft.com/en-us/seeing-ai

[4]Approximately 75 million smartwatches were sold in 2017 and this number expected to get doubled in 2018. https://www.statista.com/statistics/538237/global-smartwatch-unit-sales/

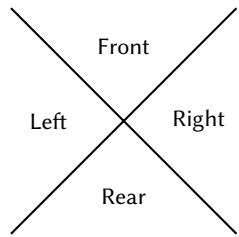

**Figure 1: We convey the positions of nearby people using four coarse zones relative to the user.**

[5]DotWatch. https://www.dotincorp.com/

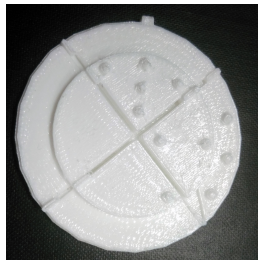

**Figure 2: A sample of our 3D printed tactile prototype, showing a situation with 5 or more people in front of the user, two people nearby to the right, and three people further off to the right.**

## DESIGN GOALS

The goal of our study is to explore prototypes to convey three pieces of information to a user with visual impairments: 1) number of people nearby; 2) their positions (e.g., compass directions relative to the user); and 3) their distances from the user. We designed three prototypes, each using a different modality of feedback: tactile, vibrotactile, and auditory. Since our goal is to provide privacy feedback, we prioritized *discreetness*, which makes wrist-worn devices attractive. To convey the position of people in the surroundings, we adopt a model in which the user's nearby space is divided into four regions: front, right, left, and rear (Figure 1). For the number of people, we provide higher fidelity (exact) counts if the number of surrounding people is few, and approximate counts for large numbers of people. We considered two levels of distance, 'near' and 'far,' without defining an explicit distance threshold as it may vary by user and situation. We anticipated that users may also want to be specifically notified if the device detects certain situations, e.g., a person coming 'too close.'

### Three Prototypes

*3D Printed Model for Tactile Feedback.* Our first prototype is a circular watch that conveys information through touch. While there are Braille smartwatches[5], we are not aware of an existing platform to easily test our prototype; since we only needed to explore the feedback mechanism, we created a 3D-printed model (Figure 2) to provide a realistic simulation of the touch experience.As shown in Figure 2, the face of the watch was partitioned into four directions (front, left, right, rear) with raised marks. To identify the front zone we included a small bump on the front edge (visible at top of figure). A raised inner circular region represented the "near" area closest to the user. Each of these 8 different zones could have between 0–5 bumps to indicate the number of people in the region, where five indicated five or more. An alternative possibility would have been to use Braille numerals, but we avoided this since Braille literacy is rapidly decreasing [8].

*Smartwatch App for Vibrations.* Our second prototype provides feedback using vibrations, and we implemented it as an app for an Android Smartwatch. In our prototype, the user requests information for a particular zone by swiping from the center of the watch towards the zone of interest. A short vibration confirms that a swipe has been detected. After a brief pause, the watch gives one long vibration to indicate that between one to five people are in that direction, two long vibrations to indicate more than five, and no vibration if there is no one. This prototype does not distinguish between near and far people, but does have an "active alerting" feature that vibrates many (10) times to indicate an unsafe situation (e.g., if someone becomes very close). We call this feature feature "active" because it continuously monitors the environment and immediately alerts the user of a situation, instead of waiting for the user to request information (through a swipe).

Table 1: Demographic information for our study participants

| ID | Age | Sex | Impairment |
| --- | --- | --- | --- |
| P1 | 31-40 | F | Totally Blind (light perception) |
| P2 | 31-40 | M | Totally Blind (light perception) |
| P3 | 31-40 | M | Low Vision |
| P4 | 41-50 | M | Totally Blind (light perception) |
| P5 | 41-50 | F | Totally Blind |
| P6 | 41-50 | F | Totally Blind |
| P7 | 51-60 | F | Totally Blind |
| P8 | 51-60 | F | Vision and Hearing Impaired |
| P9 | 51-60 | M | Totally Blind |
| P10 | 51-60 | F | Totally Blind |
| P11 | 51-60 | F | Low Vision |
| P12 | 61-70 | M | Totally Blind (light perception) |
| P13 | 61-70 | M | Totally Blind |
| P14 | 70+ | F | Totally Blind |

Table 2: Technology and Braille usage by our study participants

| ID | Technology | Braille |
| --- | --- | --- |
| P1 | Computer, iPhone, Braille display | Yes |
| P2 | Smartphone, Laptop, Braille display | Yes |
| P3 | iPhone, Desktop | Yes |
| P4 | iPhone, iPad, Macbook | Yes |
| P5 | iPhone, Computer, Laptop | Yes |
| P6 | Laptop, iPhone | Yes |
| P7 | Desktop, Laptop, Flip phone | Yes |
| P8 | iPhone, Computer, Braille writer | Yes |
| P9 | Computer, iPhone, Scanner | No |
| P10 | Computer, iPhone | Yes |
| P11 | Computer, Smartphone | Yes |
| P12 | Computer, Smartphone | No |
| P13 | Laptop, Notebook, Flip phone | Yes |
| P14 | Computer, iPhone | No |

*Smartphone app for Audio.* We also explored audio feedback, which we implemented using an Android smartphone app for convenience, although another device (e.g., a smartwatch) could be used in practice. We aimed to deliver feedback without extra accessories like headphones, since previous studies have shown that headphones are inconvenient for many people with visual impairments [2, 3]. To make the meaning of the feedback less obvious to bystanders, we avoided descriptive audio (e.g. full sentences). The app was designed to speak four numbers, in sequence, indicating the number of people in each direction (starting from front and then rotating clockwise). For example, "2 1 0 3" would indicate two people in front, one person on the right, no one behind, and 3 people on the left. As with our second prototype, this prototype did not attempt to convey distance information, but did have "active alerting" that generated a tone for unsafe situations (e.g., person very close).

## STUDY METHODOLOGY

To investigate the usability and other design traits of the prototypes, we conducted in-person semi-structured interviews with 14 participants (Table 1 and 2). We began each interview by asking participants about basic demographic information, including age, type of visual impairment, and technology use. We then described three particular scenarios that people with visual impairments have reported as exposing safety and privacy risks in past work [2, 3]: withdrawing cash at an ATM, waiting at a transit station, and using a device in public. After understanding and confirming our participants' concerns related to these scenarios, we presented our three prototypes in random order. We described each prototype and gave instructions on how to use it. To confirm that our participants understood how to use the prototypes, we asked them to carry out several tasks before continuing.

We then described three hypothetical scenarios to help participants conceptualize real-world use cases of the prototypes: 1) a private room in which someone is sitting behind the participant; 2) a public place (library) in which the participant is surrounded on three of four sides by different numbers of people; and 3) an ATM booth with a single other person to the right of the participant. We asked them to imagine performing a private activity (e.g., reading email or withdrawing money) in these scenarios, and presented them with prototypes that were designed to indicate each of these configurations of people. In total, participants were presented with 8 configurations (3 scenarios in audio and vibrations, 2 scenarios in tactile) to gauge the participants' understanding of the prototypes. We asked follow up questions to check if the prototypes were conveying the information adequately.

The interviews were audio recorded and later transcribed. The transcriptions were analyzed and coded using iterative coding with open coding where two researchers rated various issues (e.g., usability, learnability, advantages, problems) on the prototypes.

## FINDINGS

Our participants commented on the prototypes and provided suggestion to improve the design:

**Design Factor: Convenience**
Six participants felt convenience was an important factor. Participants disliked tactile feedback as it was cumbersome and the necessity of two hands made the prototype inconvenient. It was also inconvenient for people who could not feel the bumps easily, such as those suffering from diabetes.

**Design Factor: Discreetness**
From a privacy perspective, discreetness was more important for some participants (N=5) than speed as they did not want to draw attention to themselves.

**Usability of Tactile Prototype**
Among the three prototypes, the 3D-printed watch provided the most specific information (number of people and distance from the user). Nevertheless, our participants reported several issues with this prototype. For example, while those who knew Braille (N=3) grasped the design fairly quickly, several participants (N=5) struggled to distinguish between the boundaries and bumps. We noticed that all participants needed to use both hands to access information from this design, which made it much less convenient than we had anticipated (**Design Factor: Convenience**). Moreover two participants (P9 and P12) were not able to use the design at all because they suffer from diabetes, which limited their sensation of touch . However, some participants (P2, P3) mentioned that they can access information without drawing unwanted attention, which is an "appropriate way" to provide privacy feedback . Some participant (P11) felt this modality could help search for a private place when in an otherwise public setting.

*Suggested Improvements.* The tactile feedback in our prototype is printed with a singular material (plastic). Some participants (P2, P7) thought if the materials of the boundary regions and the bumps were different it would be faster to access both types of information. Even different textures and distinguishable heights of the boundary regions would help them to access the information quickly.

**Usability of Vibration-based Prototype**
Approximately half of the participants expressed a preference for vibration-based feedback because vibrations are discreet and easy to notice (**Design Factor: Discreetness**). This modality was less likely to be missed than the others, and participants grasped the design relatively quickly. The small initial vibration helped as an acknowledgement of a successful swipe, and the delay between vibrations were sufficient to distinguish them (P1, P6, P11). The "active alerting" of when someone got close was a favorite feature of this prototype, especially for those with both vision and hearing impairments (P1, P8). However, users needed to interact with the system through swipes, resulting in slower interaction than with the audio prototype. We also noticed that two participants (P6, and P10) struggled to find the correct orientation of the device.

*Suggested Improvements.* Several participants (P1, P8, P11) suggested several ways to improve the level of specificity. Currently, the design provides at most two long vibrations where one vibration indicates one to five people and two vibrations indicate there are more than five people. The design could provide several shorter vibrations to indicate exact the number of people (up to five) and one longer vibration if there are five or more people in the corresponding zone. The distance can also be incorporated by changing the intensity of the vibrations. If the person comes closer then the device can provide longer vibrations and if the person is going in the opposite direction or farther away from

**Design Factor: Speed of Conveying Information**

For most participants (N=7), the speed of conveying privacy feedback was the most important factor. Most participants mentioned that a slower method of interaction would limit the usefulness of the prototypes in protecting privacy and safety. The 3D-printed tactile model was least preferred because it took more time to access information, while many participants chose audio over the vibration prototype as it conveyed information faster:

*If you are wearing it to keep from getting mugged, then I don't think this (tactile prototype) will stop it. Because they are 15 feet away, this (device) does not know that they are there. If you are blind and they are not, they can cover that 15 feet in less time it would take you to read this. If you put it up and check it, by the time you even think, they are on you.* (P12)

Exact Number of People is not Important.
*If there is more than five people or there is just five people it doesn't really make any difference. It's a crowd either way. It's a group either way. It doesn't add anything to my sense of security to know that there is 18 or five, but it does to know if there is one or five.* (P11)

the user, then it could provide softer vibrations (P2, P6). Two participants (P2, P3) also suggested a quicker way to provide the vibrations; instead of vibrating in response to swipes, the watch could vibrate in any particular areas (only part of the watch can vibrate) to indicate the people's location which would be quicker than the current process.

**Usability of Audio Prototype**

People with visual impairments are generally accustomed to audio feedback, since they already use audio for other purposes. Audio was also the fastest method for conveying information (**Design Factor: Speed of Conveying Information**). Although audio is less discreet than other prototypes and sometimes difficult to notice (e.g., in noisy environments or when listening to other audio), most participants (N=7) liked audio feedback as it conveyed information rapidly, giving number of people and direction simultaneously, and was easy to perceive and interpret. The interviews identified this prototype as the most usable, and identified few issues with its current design. However, some participants reported that the order of the numbers might confuse directionally challenged people, and suggested using spatial audio instead.

*Suggested Improvements.* The audio design did not provide information about distance, and some participants reported that this would be necessary. They suggested including encoding distance into the feedback through, for example, audio volume (P2, P7), speech rate (P3, P4), or pitch (P2). A separate tone could also be used to signal someone moving closer (P6). Using spatial audio could ease the process for conveying direction (P7).

**Design Implications**

Advances in wearable camera and computer vision technologies may hold great promise for people with visual impairments, and some devices have already demonstrated the ability to continuously analyze the environment and provide descriptions (e.g., Seeing AI). Due to the complicated nature of our surroundings, however, it is still an open problem how to summarize and convey meaningful information to people with visual impairments unobtrusively. In our study, we evaluated three wrist-based methods to relay environmental information to people with visual impairments, with a specific focus on information that is important to manage privacy. Our participants discussed various design suggestions that should be considered for practical devices:

*Exact number of people is not important.* Most participants (N=7) felt a system should indicate the exact number of people within some specific range, up to a maximum around three (P1) to five (P2, P6, P8, P10); for larger group sizes, the exact number would not matter since the situation would not be considered private no matter what (P11). Three participants (P3, P7, P12) suggested that just

**Alert is Mandatory.**
*I think the alert system is a must. If I am walking across and somebody is coming at me and it alerts me I know to be careful for that person. If something does not alert me, then if you are both walking you can step onto each other. Even if you are not getting an alert you can't walk around feeling for a bump to pop up. I think the alert system is what makes this would make them work. (P5)*

knowing if at least one person is nearby would be sufficient to maintain their privacy, while two others suggested that coarser information would be sufficient, e.g., broad groups 1–5, and 5 or more people.

*Redefined privacy bubble.* Ahmed *et al.* [3] reported the concept of a "privacy bubble" that can be as large as 5–15 feet. In our study, we also presented the idea of a privacy bubble to our participants and provided information on two levels ('near' and 'far'). Participants found this to be a useful concept, not only for privacy but also to be aware of people's presence in social settings and for finding quiet places. Some participants reported the privacy bubble could be anywhere from 2–20 feet, although most reported 3–12 feet. P8 suggested the distance could be divided into risk zones, e.g., 2–4 feet for "high risk," 4–8 for "medium risk," and more than 8 feet for "low risk."

*Monitoring and alerts.* Most participants felt active monitoring and alerts was a required feature. Alerts would be particularly helpful for those who have both vision and hearing impairments since they cannot rely on hearing footsteps to know if people are approaching (P8). Beyond privacy and safety, an alert system may also simply prevent them from bumping into others (P11).

*Combining prototypes.* A majority of our participants suggested combining elements from multiple prototypes, although participants differed in their exact suggestions. For example, some participants wanted relatively little information, while others tended to want to know of smaller details, so a combination of different prototypes with configurable levels of feedback may be useful. Moreover, a combination of feedback modalities could help distribute the cognitive load on any one of a user's senses [8] — audio could provide quick initial information, for example, while slower but more detailed tactile feedback could be used to provide more specific information.

*SIIDs and Mobile Interaction.* In addition to a lot of environmental factors (e.g., lighting, noise [11, 14]) for situationally induced impairments and disabilities (SIIDs), people with visual impairments may experience SIIDs due to their privacy problems [1, 2]. Poor design can also introduce additional SIIDs [13] and a poorly designed solution could be ineffective for helping people with visual impairments. For better understanding the challenges of people with visual impairments, in this work we designed three wearable prototypes and conducted a usability study with 14 participants with visual impairments. Our study shows that the prototypes can be promising for addressing the privacy challenges, however, it can add additional SIIDs (e.g., the user may miss audio tones in a noisy environment). Our study suggests that a combination of prototypes may reduce the risk of SIIDs of these prototypes, however, we need additional explorations for a complete solution.

## CONCLUSIONS

We presented a study of three medium-fidelity prototypes to convey the number of people nearby and their relative distance using three modes of communication. Our prototypes might help people with visual impairments raise their situational awareness in social surroundings. Our study suggested the limitations of the prototypes and provided useful design implications for conveying information about people nearby (e.g., precise information is not always important and information about nearby people can be provided at different levels of granularity based on distance). In the future, we plan to develop a more refined prototype based on these findings and wish to conduct an in-situ study.